\documentclass[aps,prb,twocolumn,superscriptaddress,showkeys,showpacs]{revtex4-1}

\usepackage{graphicx}
\usepackage{dcolumn}
\usepackage{bm}
\usepackage{textgreek}
\usepackage{float}
\usepackage{fancyhdr}

\begin{document}

\preprint{}

\title{Epitaxial Growth and Electronic Structure of a Layered Zinc Pnictide \\Semiconductor, \textbeta-BaZn$_{\text{2}}$As$_{\text{2}}$}

\author{Zewen Xiao}
\author{Fan-Yong Ran}
\affiliation{Materials and Structure Laboratory, Tokyo Institute of Technology, Mail Box R3-4, 4259 Nagatsuta, Midori-ku, Yokohama 226-8503, JAPAN}
\author{Hidenori Hiramatsu}
\affiliation{Materials and Structure Laboratory, Tokyo Institute of Technology, Mail Box R3-4, 4259 Nagatsuta, Midori-ku, Yokohama 226-8503, JAPAN}
\affiliation{Materials Research Center for Element Strategy, Tokyo Institute of Technology, Mail Box S2-12, 4259 Nagatsuta, Midori-ku, Yokohama 226-8503, JAPAN}
\author{Satoru Matsuishi}
\affiliation{Materials Research Center for Element Strategy, Tokyo Institute of Technology, Mail Box S2-12, 4259 Nagatsuta, Midori-ku, Yokohama 226-8503, JAPAN}
\author{Hideo Hosono}
\affiliation{Materials and Structure Laboratory, Tokyo Institute of Technology, Mail Box R3-4, 4259 Nagatsuta, Midori-ku, Yokohama 226-8503, JAPAN}
\affiliation{Materials Research Center for Element Strategy, Tokyo Institute of Technology, Mail Box S2-12, 4259 Nagatsuta, Midori-ku, Yokohama 226-8503, JAPAN}
\affiliation{Frontier Research Center, Tokyo Institute of Technology, Mail Box S2-12, 4259 Nagatsuta, Midori-ku, Yokohama 226-8503, JAPAN}
\author{Toshio Kamiya}
\affiliation{Materials Research Center for Element Strategy, Tokyo Institute of Technology, Mail Box S2-12, 4259 Nagatsuta, Midori-ku, Yokohama 226-8503, JAPAN}
\affiliation{Frontier Research Center, Tokyo Institute of Technology, Mail Box S2-12, 4259 Nagatsuta, Midori-ku, Yokohama 226-8503, JAPAN}

\date{October 30, 2013}

\begin{abstract}
BaZn$_{\text{2}}$As$_{\text{2}}$ is expected for a good \textit{p}-type semiconductor and has two crystalline phases of an orthorhombic \textalpha\ phase and a higher-symmetry tetragonal \textbeta\ phase. Here, we report high-quality epitaxial films of the tetragonal \textbeta-BaZn$_{\text{2}}$As$_{\text{2}}$ were grown on single-crystal MgO (001) substrates by a reactive solid-phase epitaxy technique. Out-of-plane and in-plane epitaxial relationships between the film and the substrate were BaZn$_{\text{2}}$As$_{\text{2}}$ (00\textit{l})//MgO (001) and BaZn$_{\text{2}}$As$_{\text{2}}$ [200]//MgO [200], respectively. The full-widths at half maximum were 0.082\textsuperscript{o} for a 008 out-of-plane rocking curve and 0.342\textsuperscript{o} for a 200 in-plane rocking curve. A step-and-terrace structure was observed by atomic force microscopy. The band gap of \textbeta-BaZn$_{\text{2}}$As$_{\text{2}}$ was evaluated to be around 0.2 eV, which is much smaller than that of a family compound LaZnOAs (1.5 eV). Density functional theory calculation using the Heyd\textendash Scuseria\textendash Ernzerhof hybrid functionals supports the small band gap.
\end{abstract}


\fancypagestyle{plain}

\maketitle

\thispagestyle{plain} \lhead{Please cite this article as: Z. Xiao, et al., Thin Solid Films (2013), http://dx.doi.org/10.1016/j.tsf.2013.10.135}

\section{Introduction}
Layered mixed anion compounds such as LaCuO(S,Se,Te)\cite{ueda-apl2000,ueda-apl2001,Hiramatsu-apl2003,Hiramatsu-apl2002} and La(Fe,Ni)O(P,As)\cite{Kamihara-jacs2006,Kamihara-jacs2008,Watanabe-jssc2008} exhibit a wide variety of electronic functions such as wide band gap high-mobility \textit{p}-type semiconduction and superconductivity that make them candidates for application as functional materials. In particular, an interesting feature of these compounds is that they have a two-dimensional layered structure composed of a (Cu,Fe,Ni)\textendash(S,Se,Te,P,As) layer and a La\textendash O layer; the former forms a carrier conduction path and the latter has a wider band gap than the conduction layer and behaves like a barrier layer. These compounds with chemical formula \textit{Ln}\textit{M}O\textit{Ch} (\textit{Ln} = lanthanide, \textit{M} = transition metal, \textit{Ch} = chalcogen) are often called `1111-type' compounds. On the other hand, a similar layered crystal structure called a `122-type' one is reported, which is represented by the chemical formula \textit{Ae}\textit{M}$_{\text{2}}$\textit{Pn}$_{\text{2}}$ (\textit{Ae} = alkaline earth, \textit{Pn} = pnictogen). For the iron pnictide-based superconductors, 122-type \textit{Ae}Fe$_{\text{2}}$\textit{Pn}$_{\text{2}}$ compounds have widely been investigated due to easiness in growing high-quality epitaxial films compared to the 1111-type ones.\cite{Hiramatsu-ape2008,Katase-ssc2009,Katase-apl2011}

For semiconductor materials, we have reported that LaMnO\textit{Pn} are anti-ferromagnetic semiconductors\cite{Hanna-prb2013,Kamiya-mseb2010} and LaZnO\textit{Pn} are non-magnetic semiconductors.\cite{Kayamura-prb2007,Kayamura-tsf2008} There have, however, been only a few reports on 122-type Zn-based compounds \textit{Ae}Zn$_{\text{2}}$\textit{Pn}$_{\text{2}}$;\cite{Zhao-nc2013,Klüfers-zn1978,Hellmann-zn2007} therefore, we have focused on BaZn$_{\text{2}}$As$_{\text{2}}$ in this study. We expect a rather large band gap comparable to that of LaZnOAs (1.5 eV).\cite{Kayamura-tsf2008} Further, this materials system would be interesting also for other applications; e.g., Zhao \textit{et al}. synthesized polycrystalline (Ba,K)(Zn,Mn)$_{\text{2}}$As$_{\text{2}}$ for diluted magnetic semiconductors.\cite{Zhao-nc2013}

\begin{figure}[b]
\includegraphics[width=7.2cm]{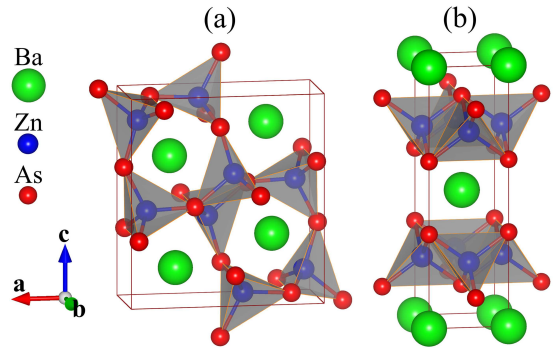}
\caption{\label{fig:1} Crystal structures of (a) orthorhombic \textalpha-BaZn$_{\text{2}}$As$_{\text{2}}$ (\textalpha-BaCu$_{\text{2}}$S$_{\text{2}}$-type, space group \textit{Pnma}), and (b) tetragonal \textbeta-BaZn$_{\text{2}}$As$_{\text{2}}$ (ThCr$_{\text{2}}$Si$_{\text{2}}$-type, space group \textit{I}4/ \textit{mmm}).}
\end{figure}

BaZn$_{\text{2}}$As$_{\text{2}}$ crystallizes into two phases; orthorhombic \textalpha-BaZn$_{\text{2}}$As$_{\text{2}}$ (\textalpha-BaCu$_{\text{2}}$S$_{\text{2}}$-type, space group  \textit{Pnma})\cite{Klüfers-zn1978} and tetragonal \textbeta-BaZn$_{\text{2}}$As$_{\text{2}}$ (ThCr$_{\text{2}}$Si$_{\text{2}}$-type, space group \textit{I}4/ \textit{mmm}),\cite{Hellmann-zn2007} as shown in Fig.~\ref{fig:1}. The \textalpha-BaZn$_{\text{2}}$As$_{\text{2}}$ consists of a 3D cage network of (ZnAs$_{\text{4/4}}$) tetrahedrons with a Ba atom encaged in each cage, which is stable at room temperature (RT) \textendash\ 850 \textsuperscript{o}C. The \textbeta-BaZn$_{\text{2}}$As$_{\text{2}}$ phase is a typical 122-type one and consists of an alternative stacking of (ZnAs$_{\text{4/4}}$) tetrahedron layers and Ba layers, which is stable at higher temperatures. Since the latter phase has a layered structure, better lateral carrier transport properties are expected in a (001)-oriented epitaxial film.

However, epitaxial growth of \textbeta-BaZn$_{\text{2}}$As$_{\text{2}}$ is difficult for the following reasons; (1) \textbeta-BaZn$_{\text{2}}$As$_{\text{2}}$ itself is of a complex structure and a meta-stable phase at RT, and bulk samples are obtained only by a rapid quenching method\cite{Hellmann-zn2007} or heavily extrinsic doping,\cite{Zhao-nc2013} (2) it is difficult to control the stoichiometric composition of BaZn$_{\text{2}}$As$_{\text{2}}$ because the Zn and As elements evaporate easily at a high temperature but such high temperature is in general necessary for epitaxial growth. For such complex compounds, we have developed a reactive solid-phase epitaxy (R-SPE) method.\cite{Hiramatsu-apl2002,Ohta-afm2003}

In this work, we report that high-quality epitaxial \textbeta-BaZn$_{\text{2}}$As$_{\text{2}}$ films were obtained by R-SPE. We also investigated the electronic structure of \textbeta-BaZn$_{\text{2}}$As$_{\text{2}}$, but found the band gap is very small (0.2 eV). Hybrid density-functional band calculation supports the small band gap.

\section{Experimental details}
\subsection{Target fabrication}
Thin films were grown by pulsed laser deposition (PLD). The BaZn$_{\text{2}}$As$_{\text{2}}$ targets were synthesized through a solid-state reaction from elementary Ba, Zn and As. Ba pieces, fresh Zn powders, and As powders, which were mixed at the stoichiometric ratio, sealed in an evacuated quartz tube, and heated at 700 \textsuperscript{o}C for 20 h. The obtained reaction product was ground, pressed into pellets (8 mm in diameter), and further pressed with a cold isostatic press (CIP) apparatus at 250 MPa. Then, the pellets were sealed in an evacuated quartz tube, reacted at 800 \textendash\ 900 \textsuperscript{o}C for 20 h, and then furnace-cooled to RT. All the target fabrication processes except for the CIP, the sealing and the heating were conducted in an argon-filled glove box. Besides, stoichiometric amorphous BaZn$_{\text{2}}$As$_{\text{2}}$ (a-BaZn$_{\text{2}}$As$_{\text{2}}$) thin films were quired for R-SPE; therefore, a composition-optimized Zn-rich target (nominal atomic ratio is Ba:Zn:As = 1: 4.0: 2.5) was fabricated by almost the same way except for a lower annealing temperature for the pellet (500 \textsuperscript{o}C) to avoid Zn precipitation.

\subsection{Thin film growth}
First, we tried to grow epitaxial \textbeta-BaZn$_{\text{2}}$As$_{\text{2}}$ films on MgO (001) substrates by a simple  PLD method using a KrF excimer laser (\textlambda\ = 248 nm; pulse duration = 20 ns; repetition frequency = 10 Hz). However, the obtained thin films incorporated little Zn when the substrate temperatures (\textit{T}$_{\text{s}}$) were $\geq$350 \textsuperscript{o}C, irrespective of varying process parameters including laser power, target composition and vacuum pressure. On the other hand, the films deposited at \textit{T}$_{\text{s}}$ \textless\ 350 \textsuperscript{o}C were always amorphous and of Zn-poor (e.g., the Ba:Zn:As atomic ratio was 24.6: 29.7: 45.7\% for RT deposition) when a stoichiometric target was used. Besides, simple post-deposition annealing of a-BaZn$_{\text{2}}$As$_{\text{2}}$ films at a high temperature led to a solid-phase epitaxial growth, but their quality was fairly poor.

\begin{figure}[t]
\includegraphics[width=8.1cm]{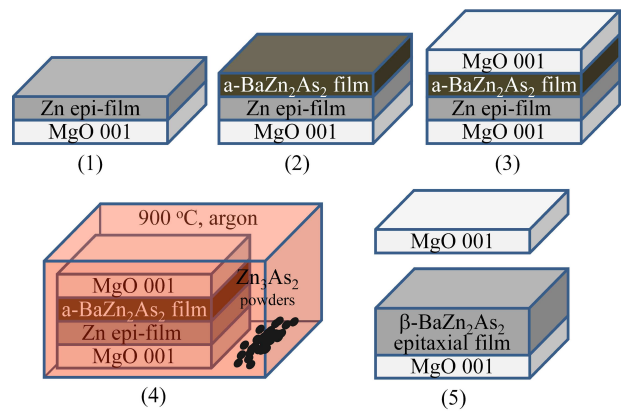}
\caption{\label{fig:2}R-SPE process. First, a sacrifice Zn thin layer was formed on a MgO (001) single crystal (1), followed by depositing an amorphous film (2). This sample was covered with a MgO (001) plate (3) and reacted in argon atmosphere at 900 \textsuperscript{o}C (4), which converted the amorphous/Zn bi-layer to a single epitaxial layer by solid-phase epitaxy from the substrate surface.}
\end{figure}

Then, we examined the R-SPE method as illustrated in Fig.~\ref{fig:2}. Firstly, a metallic Zn layer ($\sim$ 5 nm in thickness) was deposited on a single-crystal MgO (001) at RT by PLD, followed by deposition of an a-BaZn$_{\text{2}}$As$_{\text{2}}$ layer (the Ba:Zn:As atomic ratio was 18.9: 41.4: 39.7\%) at RT in the same PLD chamber using the optimized target with Ba:Zn:As = 1: 4.0: 2.5. Next, the a-BaZn$_{\text{2}}$As$_{\text{2}}$/Zn/MgO sample was covered by an MgO (001) plate, sealed in an argon-filled stainless steel tube, and annealed at 900 \textsuperscript{o}C for 30 min. A small amount of Zn$_{\text{3}}$As$_{\text{2}}$ powder was contained in the stainless steel tube to suppress the evaporation of the Zn and As elements during the post-annealing process.

\subsection{Characterization}
X-ray diffraction (XRD) patterns of the bulk samples were measured with a conventional powder XRD instrument (RINT2500, Rigaku). Crystalline quality and orientation of the grown films were examined with a high-resolution XRD (HR-XRD) instrument (Smart Lab, Rigaku). Out-of-plane (2\texttheta\textchi/\textomega\ synchronous scan) and in-plane (2\texttheta\textchi/\textphi\ synchronous scan) XRD patterns were measured to determine epitaxial relationships between the films and the MgO substrates and to calculate lattice parameters. Out-of-plane rocking curve (2\texttheta-fixed \textomega\ scan) and in-plane rocking curve (2\texttheta\textchi-fixed /\textphi\ scan) were measured to evaluate epitaxial quality of the films. Surface morphology was observed by atomic force microscope (AFM) (SPI-3700, SII). Chemical compositions were determined by X-ray fluorescence (ZSX-100e, Rigaku) with a polycrystalline sample as a reference.
Optical absorption spectra from ultraviolet to near-infrared region were recorded in air using a conventional spectrophotometer (U4100, Hitachi), and those in the mid-infrared region were performed in vacuum with a Fourier transform infrared (FTIR) spectrometer (Vertex 70v, Bruker).
\subsection{Band structure calculation}
Band structures were calculated by density functional theory (DFT) using a code VASP 5.3.3.\cite{Kresse-prb1996} We first employed Perdew\textendash Burke\textendash Ernzerhof (PBE96) general gradient approximation (GGA) functionals,\cite{Perdew-prl1996} but found that it underestimated the band gap and gave a negative value. We then examined hybrid functionals and found the Heyd\textendash Scuseria\textendash Ernzerhof (HSE) hybrid functionals\cite{Heyd-jcp2003,Heyd-jcp2006} with the standard mixing parameter of 25\% for the exact-exchange term, because we confirmed that HSE provided a reasonable result for a family compound, LaZnOAs.

\section{Results and discussion}
\begin{figure}[b]
\includegraphics[width=7.8cm]{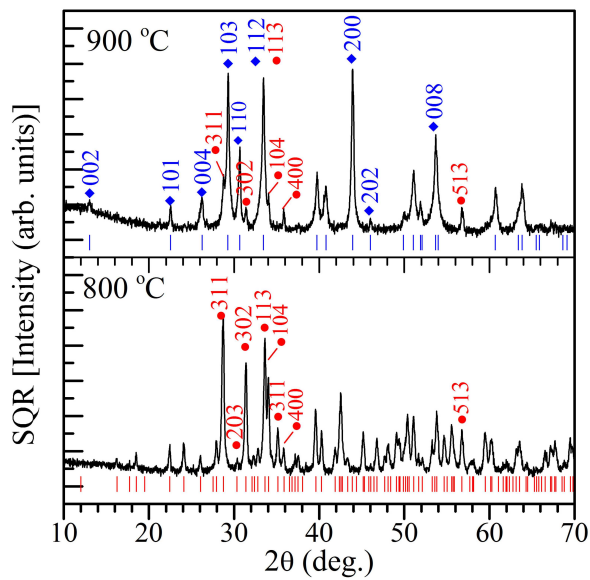}
\caption{\label{fig:3} XRD patterns of BaZn$_{\text{2}}$As$_{\text{2}}$ bulk samples synthesized at 800 and 900 \textsuperscript{o}C.}
\end{figure}

The XRD patterns of BaZn$_{\text{2}}$As$_{\text{2}}$ bulk samples are shown in Fig.~\ref{fig:3}. One synthesized at 800 \textsuperscript{o}C was of single-phase \textalpha-BaZn$_{\text{2}}$As$_{\text{2}}$ (PDF card \# 01-071-1811). That synthesized at 900 \textsuperscript{o}C was mostly the \textbeta-BaZn$_{\text{2}}$As$_{\text{2}}$ phase (\# 01-076-5124) with a small amount of the \textalpha-BaZn$_{\text{2}}$As$_{\text{2}}$ phase. This is consistent with the previous reports and it is hard to obtain a pure \textbeta-BaZn$_{\text{2}}$As$_{\text{2}}$ phase without some special methods such as the quenching method described in Ref \cite{Hellmann-zn2007}.

\begin{figure}[t]
\centering
\includegraphics[width=7.8cm]{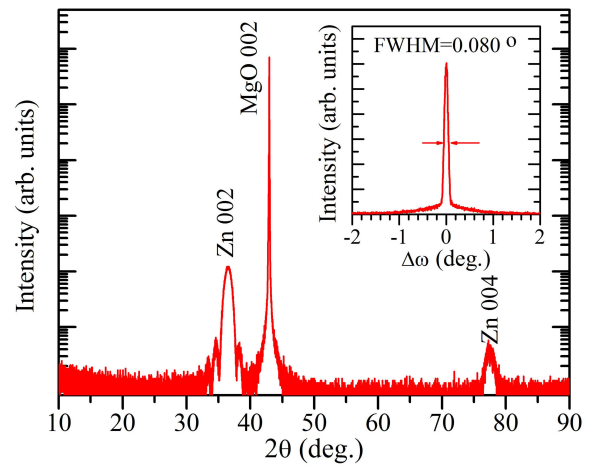}
\caption{\label{fig:4} Out-of-plane HR-XRD pattern of a-BaZn$_{\text{2}}$As$_{\text{2}}$/Zn bilayer grown on MgO (001). Inset shows out-of-plane rocking curve for the Zn 002 diffraction.}
\end{figure}

A typical out-of-plane XRD pattern of an a-BaZn$_{\text{2}}$As$_{\text{2}}$/Zn bilayer grown on a MgO (001) substrate at Rt is shown in Fig.~\ref{fig:4}. Only intense Zn 002 and 004 diffraction peaks together with a MgO 002 diffraction peak were observed, indicating the BaZn$_{\text{2}}$As$_{\text{2}}$ layer was amorphous, while the Zn layer was grown with (001) preferential orientation. The full-width at half maximum (FWHM) of the out-of-plane rocking curve for the Zn 002 diffraction (inset to Fig. 4) was 0.080\textsuperscript{o}, indicating high-quality epitaxial growth.

\begin{figure}[t]
\centering
\includegraphics[width=7.8cm]{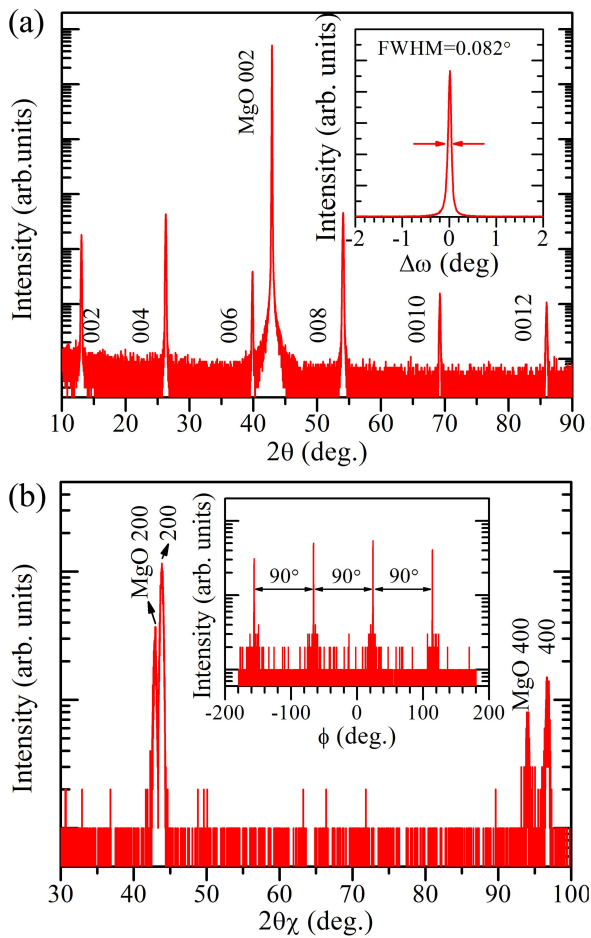}
\caption{\label{fig:5} (a) Out-of-plane and (b) in-plane HR-XRD patterns of BaZn$_{\text{2}}$As$_{\text{2}}$ thin film after annealing at 900 \textsuperscript{o}C. Inset to (a) shows out-of-plane rocking curve for the BaZn$_{\text{2}}$As$_{\text{2}}$ 008 diffraction. Inset to (b) shows in-plane rocking curve for the BaZn$_{\text{2}}$As$_{\text{2}}$ 200 diffraction.}
\end{figure}

\begin{figure}[t]
\includegraphics[width=7.2cm]{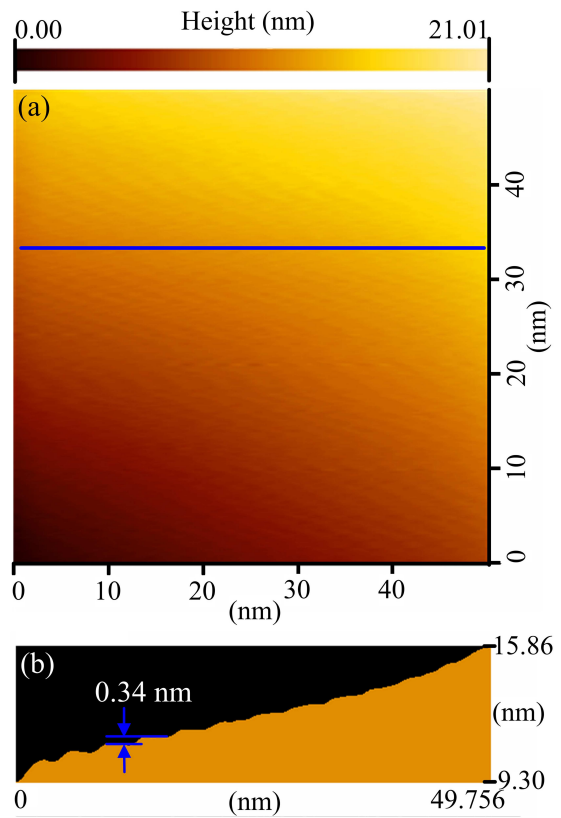}
\caption{\label{fig:6} (a) AFM image of epitaxial film annealed at 900 \textsuperscript{o}C. (b) Cross-sectional profile along the blue line shown in (a).}
\end{figure}

In Fig.~\ref{fig:5} the out-of-plane (a) and in-plane (b) HR-XRD patterns of the BaZn$_{\text{2}}$As$_{\text{2}}$ thin film fabricated by R-SPE after annealing at 900 \textsuperscript{o}C are shown.  In the out-of-plane XRD pattern, only intensive \textbeta-BaZn$_{\text{2}}$As$_{\text{2}}$ 00\textit{l} and MgO 001 diffraction peaks were observed, indicating the Zn thin layer was sacrificed, and the  \textbeta-BaZn$_{\text{2}}$As$_{\text{2}}$ phase, which is meta-stable at RT in bulk, was obtained. The ultrathin epitaxial Zn layer played as an epitaxy ‘initializer’, which is similar to an epi-ZnO layer for InGaO$_{\text{3}}$(ZnO)$_{\text{m}}$ in Ref\cite{Ohta-afm2003} and an epi-Cu layer for LaCuOS in Ref\cite{Hiramatsu-apl2002}. The FWHM of the out-of-plane rocking curve for the BaZn$_{\text{2}}$As$_{\text{2}}$ 008 diffraction (inset of Fig.~\ref{fig:5}(a)) was 0.082\textsuperscript{o}, indicating a very small tilting. In the in-plane HR-XRD pattern in Fig.~\ref{fig:5}(b), only BaZn$_{\text{2}}$As$_{\text{2}}$ 200 and 400 diffraction peaks together with MgO 200 and 400 diffraction peaks were observed. The in-plane rocking curve for 200 diffraction of BaZn$_{\text{2}}$As$_{\text{2}}$ (inset of Fig.~\ref{fig:5}(b)) exhibited a four-fold symmetry, consistent with the tetragonal symmetry of \textbeta-BaZn$_{\text{2}}$As$_{\text{2}}$ (I4/mmm). The average FWHM of the four in-plane 200 rocking curves was 0.342\textsuperscript{o}, indicating a small twisting. The lattice parameters calculated from the out-of-plane and in-plane patterns are listed in Table~\ref{tab1}. All the XRD results indicate that high-quality epitaxial \textbeta-BaZn$_{\text{2}}$As$_{\text{2}}$ thin films were obtained with the epitaxial relationship of out-of-plane BaZn$_{\text{2}}$As$_{\text{2}}$ (00l)//MgO (001) and in-plane BaZn$_{\text{2}}$As$_{\text{2}}$ (200)//MgO (200).

\begin{table}[b]
\centering
\caption{\label{tab1} Lattice parameters of epitaxial \textbeta-BaZn$_{\text{2}}$As$_{\text{2}}$ film.}
\begin{ruledtabular}
\begin{tabular}{lcdr}
\multicolumn{1}{c}{\textrm{Parameters}}&
\multicolumn{1}{c}{\textrm{Bulk or powders\footnote{PDF card: \#01-076-5124}}}&
\multicolumn{1}{c}{\textrm{Epitaxial film}}&
\textrm{Derivation (\%)}\\
\colrule
\ a (\AA) & 4.1200 & 4.1276 & +0.18\\
\ c (\AA) & 13.5780 & 13.5410 & \textendash0.30\\
V (\AA\textsuperscript{3}) & 230.4784 & 230.7332 & +0.11\\
\end{tabular}
\end{ruledtabular}
\end{table}

A typical AFM image of the epitaxial film is shown in Fig. 6.  It is confirmed that the surface was constructed by a step-and-terrace structure.  The step height was $\sim$ 0.34 nm (Fig. 6(b)), nicely agreeing with one fourth of the c-axis length c/4 (c = 1.3578 nm). The origin of this c/4 step height is not clear, but it would correspond to a single Ba layer and a single (ZnAs$_{\text{4/4}}$) layer.

\begin{figure}[t]
\centering
\includegraphics[width=8.5cm]{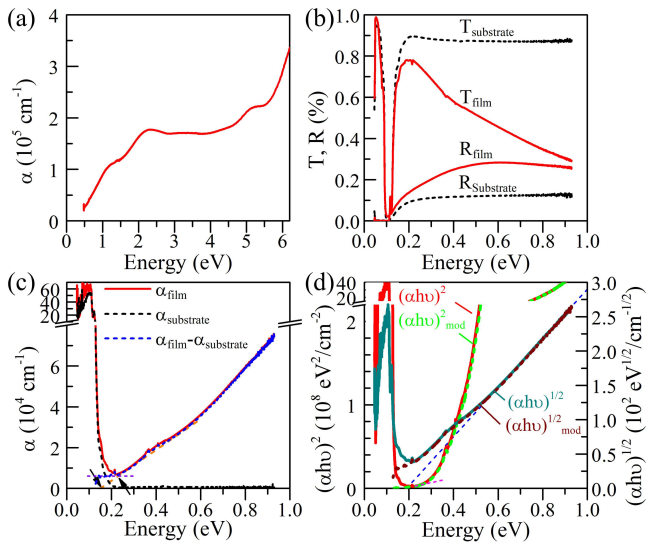}
\caption{\label{fig:7} Optical spectra of epitaxial film. (a) Absorption spectrum from ultraviolet to near-infrared region. (b) T and R spectra, (c) absorption spectra and (d) (\textit{\textalpha h\textnu})\textsuperscript{2} \textendash\ \textit{h\textnu} and (\textit{\textalpha h\textnu})\textsuperscript{1/2} \textendash\ \textit{h\textnu} plots in mid infrared region. In Figs. (b,c), the black dashed lines represent the spectra of the MgO substrate. In Figs. (c,d), the color dashed lines represent the net absorption spectra of the film after subtracting the absorption of MgO substrate.}
\end{figure}

The optical absorption spectrum of the epitaxial \textbeta-BaZn$_{\text{2}}$As$_{\text{2}}$ film measured by a conventional spectrophotometer is shown in Fig.~\ref{fig:7}(a). It exhibited a strong absorption band starting from $\sim$ 4.5 eV and a very long plateau in 2 \textendash\ 4.5 eV. An absorption edge was observed at \textless1.0 eV; but the absorption coefficient (\textalpha) didn't vanish to zero or reach a minimum until 0.5 eV, suggesting the band gap should be less than 0.5 eV. To evaluate the band gap for \textbeta-BaZn$_{\text{2}}$As$_{\text{2}}$, we further measured FTIR spectra in the lower energy region from 0.04 eV to 0.93 eV. The transmittance and reflectance spectra and the estimated absorption spectrum are shown in Figs.~\ref{fig:7}(b) and (c), respectively. A strong absorption peak is observed in 0.05 \textendash\ 0.20 eV for the film/substrate sample. It is caused by the absorption from the MgO substrate (shown by the black dashed lines in Figs.~\ref{fig:7}(b) and (c)). This absorption hinders the band gap determination more or less. By subtracting the absorption spectrum of the MgO substrate \textalpha  {substrate}, we plot the net absorption of the film \textalpha   {film}\textendash \textalpha  {substrate} by the blue dashed line, showing that an absorption edge would be at 0.23 eV. We also examined a (\textit{\textalpha h\textnu})\textsuperscript{2} \textendash\ \textit{h\textnu} plot for direct transition and a (\textit{\textalpha h\textnu})\textsuperscript{1/2} \textendash\ \textit{h\textnu} plot for indirect transition in Fig.~\ref{fig:7}(d); however, we cannot find a good linear region for these plots. Even so, both plots provide extrapolated values of 0.18 \textendash\ 0.20 eV. All the above results suggest that the band gap of \textbeta-BaZn$_{\text{2}}$As$_{\text{2}}$ should be around 0.2 eV, which is much smaller than that of a family compound LaZnAsO with the band gap of 1.5 eV.\cite{Kayamura-tsf2008}

\begin{figure}[h]
\centering
\includegraphics[width=8.6cm]{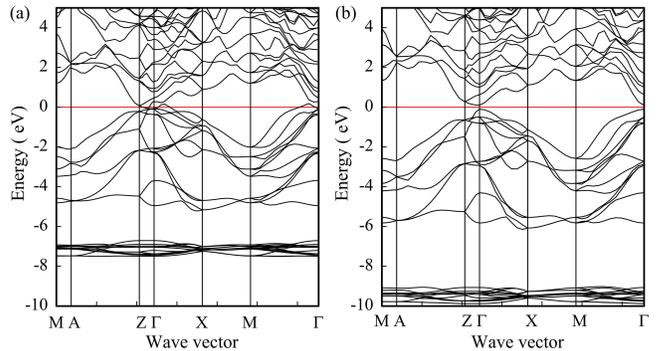}
\caption{\label{fig:8} Band structures of \textbeta-BaZn$_{\text{2}}$As$_{\text{2}}$ calculated with (a) PBE96 and (b) HSE.}
\end{figure}

We also performed band structure calculations for the tetragonal \textbeta-BaZn$_{\text{2}}$As$_{\text{2}}$. First, we employed PBE96 GGA functionals, but it gave a negative band gap structure as seen in Fig.~\ref{fig:8}(a). This is caused by a well-known band gap problem of DFT in which band gaps are in general underestimated from experimental values. Then, we examined hybrid functional calculations using HSE as shown in Fig.~\ref{fig:8}(b). It gives the band gap of 0.2 eV at the \textGamma\ point as a direct-transition one. The HSE band gap value seems to explain the optical absorption results in Fig.~\ref{fig:7}(c) and (d).

\section{Conclusions}
High-quality epitaxial \textbeta-BaZn$_{\text{2}}$As$_{\text{2}}$ films were  grown on MgO (001) substrates by the R-SPE method. The obtained phase was meta-stable high-symmetry tetragonal phase (\textit{I}4/\textit{mmm}). The epitaxial relationship was BaZn$_{\text{2}}$As$_{\text{2}}$ (00\textit{l})//MgO (001) (in-plane) and BaZn$_{\text{2}}$As$_{\text{2}}$ [200]//MgO [200] (out-of-plane). High crystalline quality was confirmed by the small tilting angle of 0.082\textsuperscript{o} and the small twisting angle of 0.342\textsuperscript{o}. The band gap of \textbeta-BaZn$_{\text{2}}$As$_{\text{2}}$ was estimated to be around 0.2 eV from the FTIR spectra and the HSE calculation.

\begin{acknowledgments}
This work was conducted under Tokodai Institute for Element Strategy (TIES) funded by MEXT Elements Strategy Initiative to Form Core Research Center.
\end{acknowledgments}

\bibliography{ThinSolidFilms-Manuscript}

\begin{thebibliography}{22}%
\makeatletter
\providecommand \@ifxundefined [1]{%
 \@ifx{#1\undefined}
}%
\providecommand \@ifnum [1]{%
 \ifnum #1\expandafter \@firstoftwo
 \else \expandafter \@secondoftwo
 \fi
}%
\providecommand \@ifx [1]{%
 \ifx #1\expandafter \@firstoftwo
 \else \expandafter \@secondoftwo
 \fi
}%
\providecommand \natexlab [1]{#1}%
\providecommand \enquote  [1]{``#1''}%
\providecommand \bibnamefont  [1]{#1}%
\providecommand \bibfnamefont [1]{#1}%
\providecommand \citenamefont [1]{#1}%
\providecommand \href@noop [0]{\@secondoftwo}%
\providecommand \href [0]{\begingroup \@sanitize@url \@href}%
\providecommand \@href[1]{\@@startlink{#1}\@@href}%
\providecommand \@@href[1]{\endgroup#1\@@endlink}%
\providecommand \@sanitize@url [0]{\catcode `\\12\catcode `\$12\catcode
  `\&12\catcode `\#12\catcode `\^12\catcode `\_12\catcode `\%12\relax}%
\providecommand \@@startlink[1]{}%
\providecommand \@@endlink[0]{}%
\providecommand \url  [0]{\begingroup\@sanitize@url \@url }%
\providecommand \@url [1]{\endgroup\@href {#1}{\urlprefix }}%
\providecommand \urlprefix  [0]{URL }%
\providecommand \Eprint [0]{\href }%
\providecommand \doibase [0]{http://dx.doi.org/}%
\providecommand \selectlanguage [0]{\@gobble}%
\providecommand \bibinfo  [0]{\@secondoftwo}%
\providecommand \bibfield  [0]{\@secondoftwo}%
\providecommand \translation [1]{[#1]}%
\providecommand \BibitemOpen [0]{}%
\providecommand \bibitemStop [0]{}%
\providecommand \bibitemNoStop [0]{.\EOS\space}%
\providecommand \EOS [0]{\spacefactor3000\relax}%
\providecommand \BibitemShut  [1]{\csname bibitem#1\endcsname}%
\let\auto@bib@innerbib\@empty
\bibitem [{\citenamefont {Ueda}\ \emph {et~al.}(2000)\citenamefont {Ueda},
  \citenamefont {Inoue}, \citenamefont {Kawazoe},\ and\ \citenamefont
  {Hosono}}]{ueda-apl2000}%
  \BibitemOpen
  \bibfield  {author} {\bibinfo {author} {\bibfnamefont {K.}~\bibnamefont
  {Ueda}}, \bibinfo {author} {\bibfnamefont {S.}~\bibnamefont {Inoue}},
  \bibinfo {author} {\bibfnamefont {H.}~\bibnamefont {Kawazoe}}, \ and\
  \bibinfo {author} {\bibfnamefont {H.}~\bibnamefont {Hosono}},\ }\href@noop {}
  {\bibfield  {journal} {\bibinfo  {journal} {Appl. Phys. Lett.}\ }\textbf
  {\bibinfo {volume} {77}},\ \bibinfo {pages} {2701} (\bibinfo {year}
  {2000})}\BibitemShut {NoStop}%
\bibitem [{\citenamefont {Ueda}\ \emph {et~al.}(2001)\citenamefont {Ueda},
  \citenamefont {Inoue},\ and\ \citenamefont {Hosono}}]{ueda-apl2001}%
  \BibitemOpen
  \bibfield  {author} {\bibinfo {author} {\bibfnamefont {K.}~\bibnamefont
  {Ueda}}, \bibinfo {author} {\bibfnamefont {S.}~\bibnamefont {Inoue}}, \ and\
  \bibinfo {author} {\bibfnamefont {H.}~\bibnamefont {Hosono}},\ }\href@noop {}
  {\bibfield  {journal} {\bibinfo  {journal} {Appl. Phys. Lett.}\ }\textbf
  {\bibinfo {volume} {78}},\ \bibinfo {pages} {2333} (\bibinfo {year}
  {2001})}\BibitemShut {NoStop}%
\bibitem [{\citenamefont {Hiramatsu}\ \emph {et~al.}(2003)\citenamefont
  {Hiramatsu}, \citenamefont {Ueda}, \citenamefont {Ohta}, \citenamefont
  {Hirano}, \citenamefont {Kamiya},\ and\ \citenamefont
  {Hosono}}]{Hiramatsu-apl2003}%
  \BibitemOpen
  \bibfield  {author} {\bibinfo {author} {\bibfnamefont {H.}~\bibnamefont
  {Hiramatsu}}, \bibinfo {author} {\bibfnamefont {K.}~\bibnamefont {Ueda}},
  \bibinfo {author} {\bibfnamefont {H.}~\bibnamefont {Ohta}}, \bibinfo {author}
  {\bibfnamefont {M.}~\bibnamefont {Hirano}}, \bibinfo {author} {\bibfnamefont
  {T.}~\bibnamefont {Kamiya}}, \ and\ \bibinfo {author} {\bibfnamefont
  {H.}~\bibnamefont {Hosono}},\ }\href@noop {} {\bibfield  {journal} {\bibinfo
  {journal} {Appl. Phys. Lett.}\ }\textbf {\bibinfo {volume} {82}},\ \bibinfo
  {pages} {1408} (\bibinfo {year} {2003})}\BibitemShut {NoStop}%
\bibitem [{\citenamefont {Hiramatsu}\ \emph {et~al.}(2002)\citenamefont
  {Hiramatsu}, \citenamefont {Ueda}, \citenamefont {Ohta}, \citenamefont
  {Orita}, \citenamefont {Hirano},\ and\ \citenamefont
  {Hosono}}]{Hiramatsu-apl2002}%
  \BibitemOpen
  \bibfield  {author} {\bibinfo {author} {\bibfnamefont {H.}~\bibnamefont
  {Hiramatsu}}, \bibinfo {author} {\bibfnamefont {K.}~\bibnamefont {Ueda}},
  \bibinfo {author} {\bibfnamefont {H.}~\bibnamefont {Ohta}}, \bibinfo {author}
  {\bibfnamefont {M.}~\bibnamefont {Orita}}, \bibinfo {author} {\bibfnamefont
  {M.}~\bibnamefont {Hirano}}, \ and\ \bibinfo {author} {\bibfnamefont
  {H.}~\bibnamefont {Hosono}},\ }\href@noop {} {\bibfield  {journal} {\bibinfo
  {journal} {Appl. Phys. Lett.}\ }\textbf {\bibinfo {volume} {81}},\ \bibinfo
  {pages} {598} (\bibinfo {year} {2002})}\BibitemShut {NoStop}%
\bibitem [{\citenamefont {Kamihara}\ \emph {et~al.}(2006)\citenamefont
  {Kamihara}, \citenamefont {Hiramatsu}, \citenamefont {Hirano}, \citenamefont
  {Kawamura}, \citenamefont {Yanagi}, \citenamefont {Kamiya},\ and\
  \citenamefont {Hosono}}]{Kamihara-jacs2006}%
  \BibitemOpen
  \bibfield  {author} {\bibinfo {author} {\bibfnamefont {Y.}~\bibnamefont
  {Kamihara}}, \bibinfo {author} {\bibfnamefont {H.}~\bibnamefont {Hiramatsu}},
  \bibinfo {author} {\bibfnamefont {M.}~\bibnamefont {Hirano}}, \bibinfo
  {author} {\bibfnamefont {R.}~\bibnamefont {Kawamura}}, \bibinfo {author}
  {\bibfnamefont {H.}~\bibnamefont {Yanagi}}, \bibinfo {author} {\bibfnamefont
  {T.}~\bibnamefont {Kamiya}}, \ and\ \bibinfo {author} {\bibfnamefont
  {H.}~\bibnamefont {Hosono}},\ }\href@noop {} {\bibfield  {journal} {\bibinfo
  {journal} {J. Am. Chem. Soc.}\ }\textbf {\bibinfo {volume} {128}},\ \bibinfo
  {pages} {10012} (\bibinfo {year} {2006})}\BibitemShut {NoStop}%
\bibitem [{\citenamefont {Kamihara}\ \emph {et~al.}(2008)\citenamefont
  {Kamihara}, \citenamefont {Watanabe}, \citenamefont {Hirano},\ and\
  \citenamefont {Hosono}}]{Kamihara-jacs2008}%
  \BibitemOpen
  \bibfield  {author} {\bibinfo {author} {\bibfnamefont {Y.}~\bibnamefont
  {Kamihara}}, \bibinfo {author} {\bibfnamefont {T.}~\bibnamefont {Watanabe}},
  \bibinfo {author} {\bibfnamefont {M.}~\bibnamefont {Hirano}}, \ and\ \bibinfo
  {author} {\bibfnamefont {H.}~\bibnamefont {Hosono}},\ }\href@noop {}
  {\bibfield  {journal} {\bibinfo  {journal} {J. Am. Chem. Soc.}\ }\textbf
  {\bibinfo {volume} {130}},\ \bibinfo {pages} {3296} (\bibinfo {year}
  {2008})}\BibitemShut {NoStop}%
\bibitem [{\citenamefont {Watanabe}\ \emph {et~al.}(2008)\citenamefont
  {Watanabe}, \citenamefont {Yanagi}, \citenamefont {Kamihara}, \citenamefont
  {Kamiya}, \citenamefont {Hirano},\ and\ \citenamefont
  {Hosono}}]{Watanabe-jssc2008}%
  \BibitemOpen
  \bibfield  {author} {\bibinfo {author} {\bibfnamefont {T.}~\bibnamefont
  {Watanabe}}, \bibinfo {author} {\bibfnamefont {H.}~\bibnamefont {Yanagi}},
  \bibinfo {author} {\bibfnamefont {Y.}~\bibnamefont {Kamihara}}, \bibinfo
  {author} {\bibfnamefont {T.}~\bibnamefont {Kamiya}}, \bibinfo {author}
  {\bibfnamefont {M.}~\bibnamefont {Hirano}}, \ and\ \bibinfo {author}
  {\bibfnamefont {H.}~\bibnamefont {Hosono}},\ }\href@noop {} {\bibfield
  {journal} {\bibinfo  {journal} {J. Solid State Chem.}\ }\textbf {\bibinfo
  {volume} {181}},\ \bibinfo {pages} {2117} (\bibinfo {year}
  {2008})}\BibitemShut {NoStop}%
\bibitem [{\citenamefont {Hiramatsu}\ \emph {et~al.}(2008)\citenamefont
  {Hiramatsu}, \citenamefont {Katase}, \citenamefont {Kamiya}, \citenamefont
  {Hirano},\ and\ \citenamefont {Hosono}}]{Hiramatsu-ape2008}%
  \BibitemOpen
  \bibfield  {author} {\bibinfo {author} {\bibfnamefont {H.}~\bibnamefont
  {Hiramatsu}}, \bibinfo {author} {\bibfnamefont {T.}~\bibnamefont {Katase}},
  \bibinfo {author} {\bibfnamefont {T.}~\bibnamefont {Kamiya}}, \bibinfo
  {author} {\bibfnamefont {M.}~\bibnamefont {Hirano}}, \ and\ \bibinfo {author}
  {\bibfnamefont {H.}~\bibnamefont {Hosono}},\ }\href@noop {} {\bibfield
  {journal} {\bibinfo  {journal} {Appl. Phys. Express}\ }\textbf {\bibinfo
  {volume} {1}},\ \bibinfo {pages} {101702} (\bibinfo {year}
  {2008})}\BibitemShut {NoStop}%
\bibitem [{\citenamefont {Katase}\ \emph {et~al.}(2009)\citenamefont {Katase},
  \citenamefont {Hiramatsu}, \citenamefont {Yanagi}, \citenamefont {Kamiya},
  \citenamefont {Hirano},\ and\ \citenamefont {Hosono}}]{Katase-ssc2009}%
  \BibitemOpen
  \bibfield  {author} {\bibinfo {author} {\bibfnamefont {T.}~\bibnamefont
  {Katase}}, \bibinfo {author} {\bibfnamefont {H.}~\bibnamefont {Hiramatsu}},
  \bibinfo {author} {\bibfnamefont {H.}~\bibnamefont {Yanagi}}, \bibinfo
  {author} {\bibfnamefont {T.}~\bibnamefont {Kamiya}}, \bibinfo {author}
  {\bibfnamefont {M.}~\bibnamefont {Hirano}}, \ and\ \bibinfo {author}
  {\bibfnamefont {H.}~\bibnamefont {Hosono}},\ }\href@noop {} {\bibfield
  {journal} {\bibinfo  {journal} {Solid State Commun.}\ }\textbf {\bibinfo
  {volume} {149}},\ \bibinfo {pages} {2121} (\bibinfo {year}
  {2009})}\BibitemShut {NoStop}%
\bibitem [{\citenamefont {Katase}\ \emph {et~al.}(2011)\citenamefont {Katase},
  \citenamefont {Hiramatsu}, \citenamefont {Matias}, \citenamefont {Sheehan},
  \citenamefont {Ishimaru}, \citenamefont {Kamiya}, \citenamefont {Tanabe},\
  and\ \citenamefont {Hosono}}]{Katase-apl2011}%
  \BibitemOpen
  \bibfield  {author} {\bibinfo {author} {\bibfnamefont {T.}~\bibnamefont
  {Katase}}, \bibinfo {author} {\bibfnamefont {H.}~\bibnamefont {Hiramatsu}},
  \bibinfo {author} {\bibfnamefont {V.}~\bibnamefont {Matias}}, \bibinfo
  {author} {\bibfnamefont {C.}~\bibnamefont {Sheehan}}, \bibinfo {author}
  {\bibfnamefont {Y.}~\bibnamefont {Ishimaru}}, \bibinfo {author}
  {\bibfnamefont {T.}~\bibnamefont {Kamiya}}, \bibinfo {author} {\bibfnamefont
  {K.}~\bibnamefont {Tanabe}}, \ and\ \bibinfo {author} {\bibfnamefont
  {H.}~\bibnamefont {Hosono}},\ }\href@noop {} {\bibfield  {journal} {\bibinfo
  {journal} {Appl. Phys. Lett.}\ }\textbf {\bibinfo {volume} {98}},\ \bibinfo
  {pages} {242510} (\bibinfo {year} {2011})}\BibitemShut {NoStop}%
\bibitem [{\citenamefont {Hanna}\ \emph {et~al.}(2013)\citenamefont {Hanna},
  \citenamefont {Matsuishi}, \citenamefont {Kodama}, \citenamefont {Otomo},
  \citenamefont {Shamoto},\ and\ \citenamefont {Hosono}}]{Hanna-prb2013}%
  \BibitemOpen
  \bibfield  {author} {\bibinfo {author} {\bibfnamefont {T.}~\bibnamefont
  {Hanna}}, \bibinfo {author} {\bibfnamefont {S.}~\bibnamefont {Matsuishi}},
  \bibinfo {author} {\bibfnamefont {K.}~\bibnamefont {Kodama}}, \bibinfo
  {author} {\bibfnamefont {T.}~\bibnamefont {Otomo}}, \bibinfo {author}
  {\bibfnamefont {S.}~\bibnamefont {Shamoto}}, \ and\ \bibinfo {author}
  {\bibfnamefont {H.}~\bibnamefont {Hosono}},\ }\href@noop {} {\bibfield
  {journal} {\bibinfo  {journal} {Phys. Rev. B}\ }\textbf {\bibinfo {volume}
  {87}},\ \bibinfo {pages} {020401} (\bibinfo {year} {2013})}\BibitemShut
  {NoStop}%
\bibitem [{\citenamefont {Kamiya}\ \emph {et~al.}(2010)\citenamefont {Kamiya},
  \citenamefont {Yanagi}, \citenamefont {Watanabe}, \citenamefont {Hirano},\
  and\ \citenamefont {Hosono}}]{Kamiya-mseb2010}%
  \BibitemOpen
  \bibfield  {author} {\bibinfo {author} {\bibfnamefont {T.}~\bibnamefont
  {Kamiya}}, \bibinfo {author} {\bibfnamefont {H.}~\bibnamefont {Yanagi}},
  \bibinfo {author} {\bibfnamefont {T.}~\bibnamefont {Watanabe}}, \bibinfo
  {author} {\bibfnamefont {M.}~\bibnamefont {Hirano}}, \ and\ \bibinfo {author}
  {\bibfnamefont {H.}~\bibnamefont {Hosono}},\ }\href@noop {} {\bibfield
  {journal} {\bibinfo  {journal} {Mater. Sci. Eng. B}\ }\textbf {\bibinfo
  {volume} {173}},\ \bibinfo {pages} {239} (\bibinfo {year}
  {2010})}\BibitemShut {NoStop}%
\bibitem [{\citenamefont {Kayamura}\ \emph {et~al.}(2007)\citenamefont
  {Kayamura}, \citenamefont {Hiramatsu}, \citenamefont {Hirano}, \citenamefont
  {Kawamura}, \citenamefont {Yanagi}, \citenamefont {Kamiya},\ and\
  \citenamefont {Hosono}}]{Kayamura-prb2007}%
  \BibitemOpen
  \bibfield  {author} {\bibinfo {author} {\bibfnamefont {K.}~\bibnamefont
  {Kayamura}}, \bibinfo {author} {\bibfnamefont {H.}~\bibnamefont {Hiramatsu}},
  \bibinfo {author} {\bibfnamefont {M.}~\bibnamefont {Hirano}}, \bibinfo
  {author} {\bibfnamefont {R.}~\bibnamefont {Kawamura}}, \bibinfo {author}
  {\bibfnamefont {H.}~\bibnamefont {Yanagi}}, \bibinfo {author} {\bibfnamefont
  {T.}~\bibnamefont {Kamiya}}, \ and\ \bibinfo {author} {\bibfnamefont
  {H.}~\bibnamefont {Hosono}},\ }\href@noop {} {\bibfield  {journal} {\bibinfo
  {journal} {Phys. Rev. B}\ }\textbf {\bibinfo {volume} {76}},\ \bibinfo
  {pages} {195325} (\bibinfo {year} {2007})}\BibitemShut {NoStop}%
\bibitem [{\citenamefont {Kayamura}\ \emph {et~al.}(2008)\citenamefont
  {Kayamura}, \citenamefont {Kawamura}, \citenamefont {Hiramatsu},
  \citenamefont {Yanagi}, \citenamefont {Hirano}, \citenamefont {Kamiya},\ and\
  \citenamefont {Hosono}}]{Kayamura-tsf2008}%
  \BibitemOpen
  \bibfield  {author} {\bibinfo {author} {\bibfnamefont {K.}~\bibnamefont
  {Kayamura}}, \bibinfo {author} {\bibfnamefont {R.}~\bibnamefont {Kawamura}},
  \bibinfo {author} {\bibfnamefont {H.}~\bibnamefont {Hiramatsu}}, \bibinfo
  {author} {\bibfnamefont {H.}~\bibnamefont {Yanagi}}, \bibinfo {author}
  {\bibfnamefont {M.}~\bibnamefont {Hirano}}, \bibinfo {author} {\bibfnamefont
  {T.}~\bibnamefont {Kamiya}}, \ and\ \bibinfo {author} {\bibfnamefont
  {H.}~\bibnamefont {Hosono}},\ }\href@noop {} {\bibfield  {journal} {\bibinfo
  {journal} {Thin Solid Films}\ }\textbf {\bibinfo {volume} {516}},\ \bibinfo
  {pages} {5800} (\bibinfo {year} {2008})}\BibitemShut {NoStop}%
\bibitem [{\citenamefont {Zhao}\ \emph {et~al.}(2013)\citenamefont {Zhao},
  \citenamefont {Deng}, \citenamefont {Wang}, \citenamefont {Han},
  \citenamefont {Zhu}, \citenamefont {Li}, \citenamefont {Liu}, \citenamefont
  {Yu}, \citenamefont {Goko}, \citenamefont {Frandsen}, \citenamefont {Liu},
  \citenamefont {Ning}, \citenamefont {Uemura}, \citenamefont {Dabkowska},
  \citenamefont {Luke}, \citenamefont {Luetkens}, \citenamefont {Morenzoni},
  \citenamefont {Dunsiger}, \citenamefont {Senyshyn}, \citenamefont {Böni},\
  and\ \citenamefont {Jin}}]{Zhao-nc2013}%
  \BibitemOpen
  \bibfield  {author} {\bibinfo {author} {\bibfnamefont {K.}~\bibnamefont
  {Zhao}}, \bibinfo {author} {\bibfnamefont {Z.}~\bibnamefont {Deng}}, \bibinfo
  {author} {\bibfnamefont {X.}~\bibnamefont {Wang}}, \bibinfo {author}
  {\bibfnamefont {W.}~\bibnamefont {Han}}, \bibinfo {author} {\bibfnamefont
  {J.}~\bibnamefont {Zhu}}, \bibinfo {author} {\bibfnamefont {X.}~\bibnamefont
  {Li}}, \bibinfo {author} {\bibfnamefont {Q.}~\bibnamefont {Liu}}, \bibinfo
  {author} {\bibfnamefont {R.}~\bibnamefont {Yu}}, \bibinfo {author}
  {\bibfnamefont {T.}~\bibnamefont {Goko}}, \bibinfo {author} {\bibfnamefont
  {B.}~\bibnamefont {Frandsen}}, \bibinfo {author} {\bibfnamefont
  {L.}~\bibnamefont {Liu}}, \bibinfo {author} {\bibfnamefont {F.}~\bibnamefont
  {Ning}}, \bibinfo {author} {\bibfnamefont {Y.}~\bibnamefont {Uemura}},
  \bibinfo {author} {\bibfnamefont {H.}~\bibnamefont {Dabkowska}}, \bibinfo
  {author} {\bibfnamefont {G.}~\bibnamefont {Luke}}, \bibinfo {author}
  {\bibfnamefont {H.}~\bibnamefont {Luetkens}}, \bibinfo {author}
  {\bibfnamefont {E.}~\bibnamefont {Morenzoni}}, \bibinfo {author}
  {\bibfnamefont {S.}~\bibnamefont {Dunsiger}}, \bibinfo {author}
  {\bibfnamefont {A.}~\bibnamefont {Senyshyn}}, \bibinfo {author}
  {\bibfnamefont {P.}~\bibnamefont {Böni}}, \ and\ \bibinfo {author}
  {\bibfnamefont {C.}~\bibnamefont {Jin}},\ }\href@noop {} {\bibfield
  {journal} {\bibinfo  {journal} {Nat. Commun.}\ }\textbf {\bibinfo {volume}
  {4}},\ \bibinfo {pages} {1442} (\bibinfo {year} {2013})}\BibitemShut
  {NoStop}%
\bibitem [{\citenamefont {Klüfers}\ and\ \citenamefont
  {Mewis}(1978)}]{Klüfers-zn1978}%
  \BibitemOpen
  \bibfield  {author} {\bibinfo {author} {\bibfnamefont {P.}~\bibnamefont
  {Klüfers}}\ and\ \bibinfo {author} {\bibfnamefont {A.}~\bibnamefont
  {Mewis}},\ }\href@noop {} {\bibfield  {journal} {\bibinfo  {journal} {Z.
  Naturforsch.}\ }\textbf {\bibinfo {volume} {33b}},\ \bibinfo {pages} {151}
  (\bibinfo {year} {1978})}\BibitemShut {NoStop}%
\bibitem [{\citenamefont {Hellmann}\ \emph {et~al.}(2007)\citenamefont
  {Hellmann}, \citenamefont {Löhken}, \citenamefont {Wurth},\ and\
  \citenamefont {Mewis}}]{Hellmann-zn2007}%
  \BibitemOpen
  \bibfield  {author} {\bibinfo {author} {\bibfnamefont {A.}~\bibnamefont
  {Hellmann}}, \bibinfo {author} {\bibfnamefont {A.}~\bibnamefont {Löhken}},
  \bibinfo {author} {\bibfnamefont {A.}~\bibnamefont {Wurth}}, \ and\ \bibinfo
  {author} {\bibfnamefont {A.}~\bibnamefont {Mewis}},\ }\href@noop {}
  {\bibfield  {journal} {\bibinfo  {journal} {Z. Naturforsch..}\ }\textbf
  {\bibinfo {volume} {62b}},\ \bibinfo {pages} {155} (\bibinfo {year}
  {2007})}\BibitemShut {NoStop}%
\bibitem [{\citenamefont {Ohta}\ \emph {et~al.}(2003)\citenamefont {Ohta},
  \citenamefont {Nomura}, \citenamefont {Orita}, \citenamefont {Hirano},
  \citenamefont {Ueda}, \citenamefont {Suzuki}, \citenamefont {Ikuhara},\ and\
  \citenamefont {Hosono}}]{Ohta-afm2003}%
  \BibitemOpen
  \bibfield  {author} {\bibinfo {author} {\bibfnamefont {H.}~\bibnamefont
  {Ohta}}, \bibinfo {author} {\bibfnamefont {K.}~\bibnamefont {Nomura}},
  \bibinfo {author} {\bibfnamefont {M.}~\bibnamefont {Orita}}, \bibinfo
  {author} {\bibfnamefont {M.}~\bibnamefont {Hirano}}, \bibinfo {author}
  {\bibfnamefont {K.}~\bibnamefont {Ueda}}, \bibinfo {author} {\bibfnamefont
  {T.}~\bibnamefont {Suzuki}}, \bibinfo {author} {\bibfnamefont
  {Y.}~\bibnamefont {Ikuhara}}, \ and\ \bibinfo {author} {\bibfnamefont
  {H.}~\bibnamefont {Hosono}},\ }\href@noop {} {\bibfield  {journal} {\bibinfo
  {journal} {Adv. Funct. Mater.}\ }\textbf {\bibinfo {volume} {13}},\ \bibinfo
  {pages} {139} (\bibinfo {year} {2003})}\BibitemShut {NoStop}%
\bibitem [{\citenamefont {Kresse}\ and\ \citenamefont
  {Furthmüller}(1996)}]{Kresse-prb1996}%
  \BibitemOpen
  \bibfield  {author} {\bibinfo {author} {\bibfnamefont {G.}~\bibnamefont
  {Kresse}}\ and\ \bibinfo {author} {\bibfnamefont {J.}~\bibnamefont
  {Furthmüller}},\ }\href@noop {} {\bibfield  {journal} {\bibinfo  {journal}
  {Phys. Rev. B}\ }\textbf {\bibinfo {volume} {54}},\ \bibinfo {pages} {11169}
  (\bibinfo {year} {1996})}\BibitemShut {NoStop}%
\bibitem [{\citenamefont {Perdew}\ \emph {et~al.}(1996)\citenamefont {Perdew},
  \citenamefont {Burke},\ and\ \citenamefont {Ernzerhof}}]{Perdew-prl1996}%
  \BibitemOpen
  \bibfield  {author} {\bibinfo {author} {\bibfnamefont {J.}~\bibnamefont
  {Perdew}}, \bibinfo {author} {\bibfnamefont {K.}~\bibnamefont {Burke}}, \
  and\ \bibinfo {author} {\bibfnamefont {M.}~\bibnamefont {Ernzerhof}},\
  }\href@noop {} {\bibfield  {journal} {\bibinfo  {journal} {Phys. Rev. Lett.}\
  }\textbf {\bibinfo {volume} {77}},\ \bibinfo {pages} {3865} (\bibinfo {year}
  {1996})}\BibitemShut {NoStop}%
\bibitem [{\citenamefont {Heyd}\ \emph {et~al.}(2003)\citenamefont {Heyd},
  \citenamefont {Scuseria},\ and\ \citenamefont {Ernzerhof}}]{Heyd-jcp2003}%
  \BibitemOpen
  \bibfield  {author} {\bibinfo {author} {\bibfnamefont {J.}~\bibnamefont
  {Heyd}}, \bibinfo {author} {\bibfnamefont {G.}~\bibnamefont {Scuseria}}, \
  and\ \bibinfo {author} {\bibfnamefont {M.}~\bibnamefont {Ernzerhof}},\
  }\href@noop {} {\bibfield  {journal} {\bibinfo  {journal} {J. Chem. Phys.}\
  }\textbf {\bibinfo {volume} {118}},\ \bibinfo {pages} {8207} (\bibinfo {year}
  {2003})}\BibitemShut {NoStop}%
\bibitem [{\citenamefont {Heyd}\ \emph {et~al.}(2006)\citenamefont {Heyd},
  \citenamefont {Scuseria},\ and\ \citenamefont {Ernzerhof}}]{Heyd-jcp2006}%
  \BibitemOpen
  \bibfield  {author} {\bibinfo {author} {\bibfnamefont {J.}~\bibnamefont
  {Heyd}}, \bibinfo {author} {\bibfnamefont {G.}~\bibnamefont {Scuseria}}, \
  and\ \bibinfo {author} {\bibfnamefont {M.}~\bibnamefont {Ernzerhof}},\
  }\href@noop {} {\bibfield  {journal} {\bibinfo  {journal} {J. Chem. Phys.}\
  }\textbf {\bibinfo {volume} {124}},\ \bibinfo {pages} {219906} (\bibinfo
  {year} {2006})}\BibitemShut {NoStop}%
\end{thebibliography}%

\end{document}